\documentclass[prl,superscriptaddress,twocolumn,showpacs,preprintnumbers,%
amsmath,amssymb]{revtex4}

\usepackage{graphicx}% Include figure files
\usepackage{dcolumn}% Align table columns on decimal point
\usepackage{bm}% bold math

\begin{document}   
%\title{}
\title{Momentum dependence of the electron-phonon coupling and 
self-energy effects in YBa$_2$Cu$_3$O$_7$ within the local 
density approximation}

\author{Rolf Heid}
\affiliation{Forschungszentrum Karlsruhe, Institut f\"ur Festk\"orperphysik,
P.O.B. 3640, D-76021 Karlsruhe, Germany}
\author{Klaus-Peter Bohnen}
\affiliation{Forschungszentrum Karlsruhe, Institut f\"ur Festk\"orperphysik,
P.O.B. 3640, D-76021 Karlsruhe, Germany}
\author{Roland Zeyher}
\affiliation{Max-Planck-Institut f\"ur Festk\"orperforschung,
             Heisenbergstrasse 1, D-70569 Stuttgart, Germany}
\author{Dirk Manske}
\affiliation{Max-Planck-Institut f\"ur Festk\"orperforschung,
             Heisenbergstrasse 1, D-70569 Stuttgart, Germany}

\date{\today}

\begin{abstract}
Using the local density approximation (LDA) and a realistic phonon
spectrum we determine the momentum and frequency
dependence of $\alpha^2F({\bf k},\omega)$ in YBa$_2$Cu$_3$O$_7$ 
for the bonding, antibonding, and chain band. The resulting self-energy
$\Sigma$ is rather small near the Fermi surface. For instance,
for the antibonding band the maximum of $Re \Sigma$ as a function of
frequency is about 7 meV at the nodal point in the normal state and the 
ratio of bare and renormalized Fermi velocities 1.18. These values are a 
factor 3-5 too small compared to the experiment showing that only a small 
part of $\Sigma$ can be attributed to phonons. Furthermore, the frequency 
dependence of the renormalization factor $Z({\bf k},\omega)$ 
is smooth and has no anomalies at the observed kink frequencies which means 
that phonons cannot produce well-pronounced kinks in stoichiometric 
YBa$_2$Cu$_3$O$_7$, at least, within the LDA.
\end{abstract}


\maketitle

%\section{\label{sec:intro} Introduction}

Angle resolved photoemission (ARPES) experiments in high-T$_c$ cuprates 
show that electrons near the Fermi surface interact strongly 
with bosonic excitations causing a kink in their dispersion at rather 
well-defined energies\cite{Bogdanov,Lanzara,Gromko}. 
The most popular candidates for the bosonic excitations are spin
fluctuations\cite{Manske} and phonons\cite{Lanzara}. 
Plausible arguments have been put forward for either 
of the two choices but no consensus has been 
reached up to now which of these
two possibilities actually is realized in the cuprates.
Recently, an explanation of these effects entirely in terms of phonons 
has been proposed identifying the 40 and 70 meV kink energies with
the buckling and the breathing phonon modes\cite{Devereaux}. It was also 
argued that the electron-phonon (ep) coupling in these systems
is strongly anisotropic with respect to momenta and 
phonon branches and that the coupling to the breathing  and the 
buckling phonon modes takes place at different points in the Brillouin 
zone\cite{Devereaux}, which would provide a natural explanation of
several observations.

Assuming that the LDA for stoichiometric YBa$_2$Cu$_3$O$_7$ represents a
reasonable first approximation for the electron-phonon coupling in doped
cuprates it is the aim of this Letter to find out whether an 
explanation of the kink phenomenon entirely in terms of phonons is possible.
To this end we generalize previous calculations\cite{Bohnen} for the 
Eliashberg function where both momenta were averaged 
over the Fermi surface to the case where the average is done only with 
respect to one momentum. Multiplying with $2/\omega$ 
and considering a general electronic band $\nu$ the 
resulting function $\lambda_\nu({\bf k},\omega)$ can be used to study the
anisotropy of the electron-phonon interaction as a function of 
the momentum and frequency using realistic phonon branches determined from
first-principles. Our approach omits possible effects due to strong electronic
correlations beyond the LDA, a problem which presently is
unsettled. This shortcoming seems not to be very serious for our work 
and conclusions because several calculations suggest\cite{Zeyher1,Huang} that
strong electronic correlations tend to decrease rather to increase
the ep-interaction in $\bf q$-integrated quantities such as the
self-energy.

The momentum and frequency dependent coupling function 
$\lambda_{\nu}({\bf k},\omega)$ is defined by\cite{Mahan}
\begin{eqnarray}
\lambda_{\nu}({\bf k},\omega) = \frac{2}{\omega} 
\sum_{{\bf q},j,\mu} |g_j({\bf k}\nu,{\bf k}+{\bf q}\mu)|^2
\cdot \nonumber \\
\delta(\omega-\omega_{{\bf q}j})
\delta(\epsilon_{{\bf k}+{\bf q}\mu}).
\label{lambda}
\end{eqnarray}
$g_j({\bf k}\nu,{\bf k}+{\bf q}\mu)$ denotes the amplitude for a
transition from the electronic state with momentum $\bf k$ and band
index $\nu$ to the state with momentum ${\bf k}+{\bf q}$ and band index $\mu$
creating (annihilating) a phonon with branch label $j$ and 
momentum ${\bf q}$ $(-{\bf q})$. $\omega_{{\bf q}j}$ and 
$\epsilon_{{\bf k}\nu}$ are the eigenenergies of
the phonons and electrons, respectively, the latter are measured relative
to the chemical potential. Additional calculations suggest that $\lambda$
can be considered to be diagonal in the band indices.
A direct evaluation of Eq.(\ref{lambda}) is difficult because 
high numerical costs restricted the calculation of the matrix elements $g$
to rather coarse $\bf q$-meshes.
Large fluctuations due to a small number
of terms in Eq.(\ref{lambda}) can be avoided, however, by introducing 
Fermi-surface harmonics \cite{Allen}. 
In this approach, the ${\bf k}$-dependence of $\lambda_\nu({\bf
k},\omega)$ is expanded in a basis set of functions $W$ formed by products
of an even power of electron velocities
$v_\alpha({\bf k}\nu) =
{\partial \epsilon_{{\bf k}\nu}}/
{\partial {\bf k}_\alpha}$ where $\alpha$ is a cartesian index. We
explicitly write
\begin{equation}
\lambda_\nu({\bf k},\omega) = \sum_{\bf n} W({\bf n},{\bf k}\nu) 
\Lambda_\nu({\bf n},\omega),
\label{lam1}
\end{equation}
\begin{equation}
W({\bf n},{\bf k}\nu) = v_x^{2n_x}({\bf k}\nu)
v_y^{2n_y}({\bf k}\nu)v_z^{2n_z}({\bf k}\nu),
\label{lam2}
\end{equation}
with ${\bf n}=(n_x,n_y,n_z)$. 
Defining the Fermi surface average of a variable $X({\bf k})$ by
\begin{equation}
{\langle \langle X \rangle \rangle}_\nu =
\sum_{\bf k} \delta(\epsilon_{{\bf k}\nu})X({\bf k})/
\sum_{\bf k} \delta(\epsilon_{{\bf k}\nu}),
\label{Norm}
\end{equation}
we define
\begin{equation}
\lambda_\nu({\bf n},\omega) = {\langle \langle W({\bf n},{\bf k}\nu)
\lambda_\nu({\bf k},\omega) \rangle \rangle}_\nu.
\label{lam3}
\end{equation}
Inserting Eq.(\ref{lam1}) into Eq.(\ref{lam3}) one can 
solve for the coefficients $\Lambda_\nu({{\bf n},\omega})$ and one obtains,
\begin{equation}
\lambda_\nu({\bf k},\omega) = \sum_{\bf n} W({\bf n},{\bf k}\nu)
\sum_{\bf m} S_\nu^{-1}({\bf n},{\bf m}) \lambda_\nu({\bf m},\omega),
\label{lam4}
\end{equation}
with $$ S_\nu({\bf n},{\bf m}) = {\langle \langle W({\bf n},{\bf k}\nu)
W({\bf m},{\bf k}\nu) \rangle \rangle}_\nu.$$
Calculation of the quantities $\lambda_\nu({\bf n},\omega)$ involves
Fermi-surface averages of the squared matrix elements $g$ weighted by the basis
functions $W$, which can be easily performed within LDA.
Once  $\lambda_\nu({\bf n},\omega)$ are known, $\lambda_\nu({\bf
k},\omega)$ can be obtained from Eq.(\ref{lam4}).
Typically 10 functions $W({\bf n},{\bf k}\nu)$ were used in the 
following calculations. The sum over electron momenta in
Eq.(\ref{Norm}) was carried out on a dense 36x36x4-mesh.
For the momenta $\bf q$ we used a 12x12x4-mesh.
$\delta$-functions in Eqs.(\ref{lambda}) and (\ref{Norm}) involving
electronic energies were approximated by
a Gaussian with a width of 0.2 eV. Additional calculations using
more basis functions and a smaller width for the Gaussian showed
that our results are reasonably stable against larger basis sets or a
refined representation of the $\delta$-functions.

\begin{figure}
\includegraphics[width=8.5cm,bb= 36 828 563 246,clip]{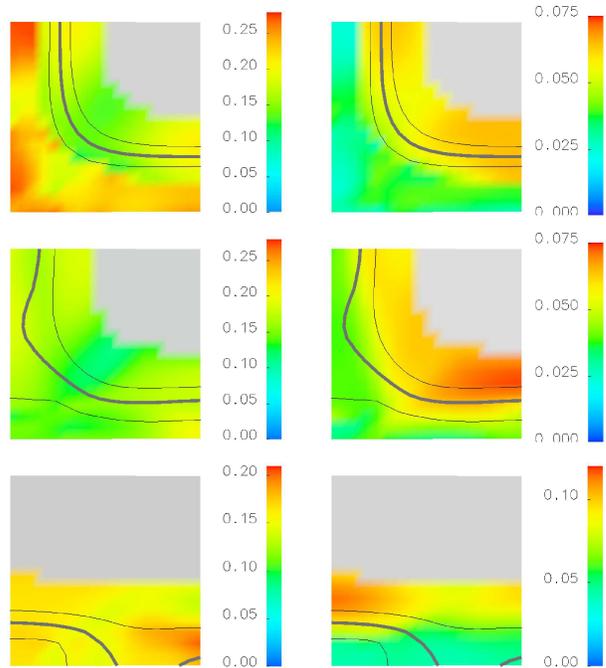}
\vspace*{67ex}
\caption{\label{fig:1}
%(Color online)
Coupling function $\lambda_\nu({\bf k},\omega)$ integrated over $\omega$ 
between 0 and 50 meV (left diagrams) and $\omega$ larger than 50 meV 
(right diagrams). Shown are $xy$-cuts through the irreducible Brillouin
zone ($0\le k_{x/y} \le \pi$) for $k_z=0.125$. The upper plots refer to 
the bonding, the middle
ones
to the antibonding, and the lower ones to the chain band. 
The latter two are hybridized in the ($\pi$,0) direction.
The grey areas represent states with energies far away ($ \ge 0.8$\,eV)
from the Fermi energy.
}
\end{figure}

Fig.\ref{fig:1} shows $\lambda_{\nu}({\bf k},\omega)$ integrated
over the frequency from 0 to 50 meV (left column) and from 50 to
80 meV (right column). The diagrams in the first row refer to the 
bonding, those in the second row to the antibonding, and those in the
third row to the chain band. 
The latter two are partly hybridized in the ($\pi$,0) direction.
The diagrams were calculated for $k_z=0.125$, but the corresponding ones
for $k_z=0.375$ look very similar, i.e., the diagrams are
practically independent of the momentum perpendicular to the planes.
The thick and thin black lines correspond to lines with 
$\epsilon_{{\bf k}\nu}=0$ and $=\pm0.2 eV$, respectively. 
As the above numerical procedure incorporates only quantities in the
vicinity of the Fermi surface, the calculated values for $\lambda_\nu({\bf
k},\omega)$ are reliable in the region between the thin black
lines whereas larger errors may be expected for momenta far away from
the Fermi surface.
%%%%%%%%%%%%%%%%%%%%%%%%%%%%%%%% FIG. 1 %%%%%%%%%%%%%%%%%%%%%%%%%%%%%%%%%%%%
%%%%%%%%%%%%%%%%%%%%%%%%%%%%%%%%%%%%%%%%%%%%%%%%%%%%%%%%%%%%%%%%%%%%%%%%%%%%

The diagrams in the first row of 
Fig.\ref{fig:1} show that the coupling functions associated with the
bonding band have approximately tetragonal symmetry but that they vary 
along the Fermi line: Their minima and maxima occur at the nodal and
antinodal points, respectively, leading to a variation of about a factor
1.5 and 1.3 for phonons below and above 50 meV, respectively.
Whereas a momentum-dependent $\lambda$ would be in line with the scenario of
Ref.\cite{Devereaux} there are serious discrepancies:
The coupling to high-frequency phonons,
for instance to the breathing and half-breathing modes, should
show a maximum in nodal direction whereas we find that it has there a minimum  
similar as in the case of low-frequency modes. 
In other words, the coupling 
to high- and low-frequency modes at the nodal and antinodal points is very
similar which means that electronic self-energies due to phonons 
at these two points 
will not select predominantly low- or predominantly 
high-frequency phonons. 
Furthermore, the variation of $\lambda$
along the Fermi line is rather small in our case, i.e., $\lambda$
cannot be considered to be very anisotropic.
The diagrams in the second row of Fig.\ref{fig:1}
describe $\lambda$ associated with the antibonding band, 
except for a part close to $(\pi,0)$, where an exchange of character
with the chain band occurs (lower diagrams of Fig.\ref{fig:1}).
However, the antibonding states in the antinodal direction along
$k_y\approx\pi$ and in the nodal direction are not affected by this
hybridization. The momentum dependence of $\lambda$ associated with the 
antibonding band resembles that of the bonding band.
%, in particular, the
%minimum is always at or near the nodal direction and the variation 
%along the Fermi line is less than a factor two.
The coupling of electrons 
near the Fermi line to low- and to high-frequency 
phonons is again rather similar as a function of ${\bf k}$, the absolute
values for $\lambda$ are roughly larger by a factor 3 for low- than for
high-frequency phonons, similar as in the case of the bonding band.   
The same trend is found for the chain band, with an anisotropy of the
order of two or smaller. Using a 4x4x2-mesh for $\bf q$ yields qualitatively 
similar results for $\lambda$ as the 12x12x4-mesh. Quantitatively,
the difference between bond and antibonding bands as well as their anisotropies
are more pronounced in this case, in particular, for the bonding band. 
The present inversion procedure leads in this case also to unphysical 
small negative values for $\lambda$ in a small region away from the Fermi line
for the bonding band which does not occur for the employed finer
12x12x4-mesh.
%%%%%%%%%%%%%%%%%%%%%%%%%%%%%%%% FIG. 2 %%%%%%%%%%%%%%%%%%%%%%%%%%%%%%%%%%%%
\begin{figure} 
\vspace*{-1.4ex}
\includegraphics[angle=0,width=6.2cm]{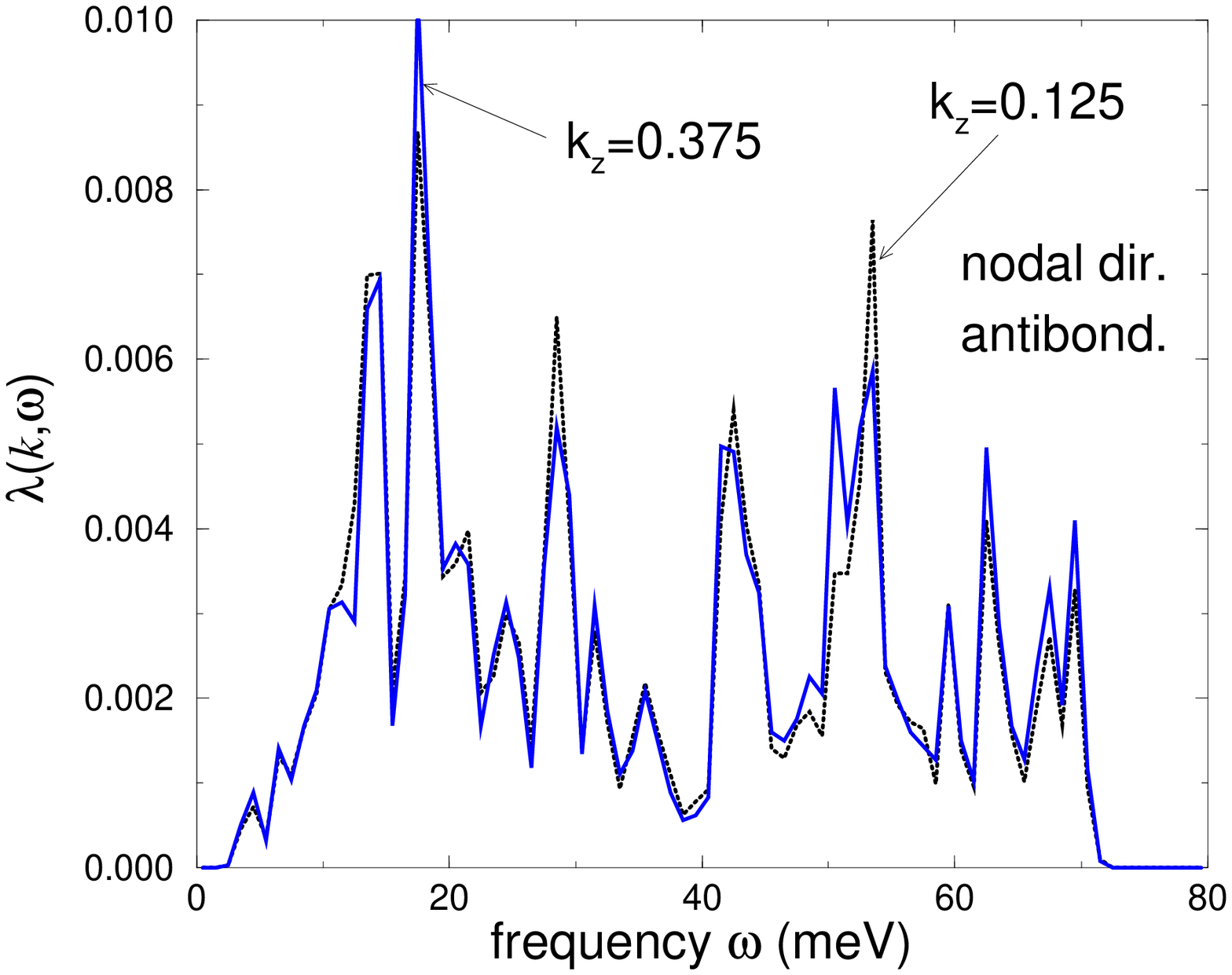}
\vspace*{-5ex}
\centerline{\includegraphics[angle=0,width=6.2cm]{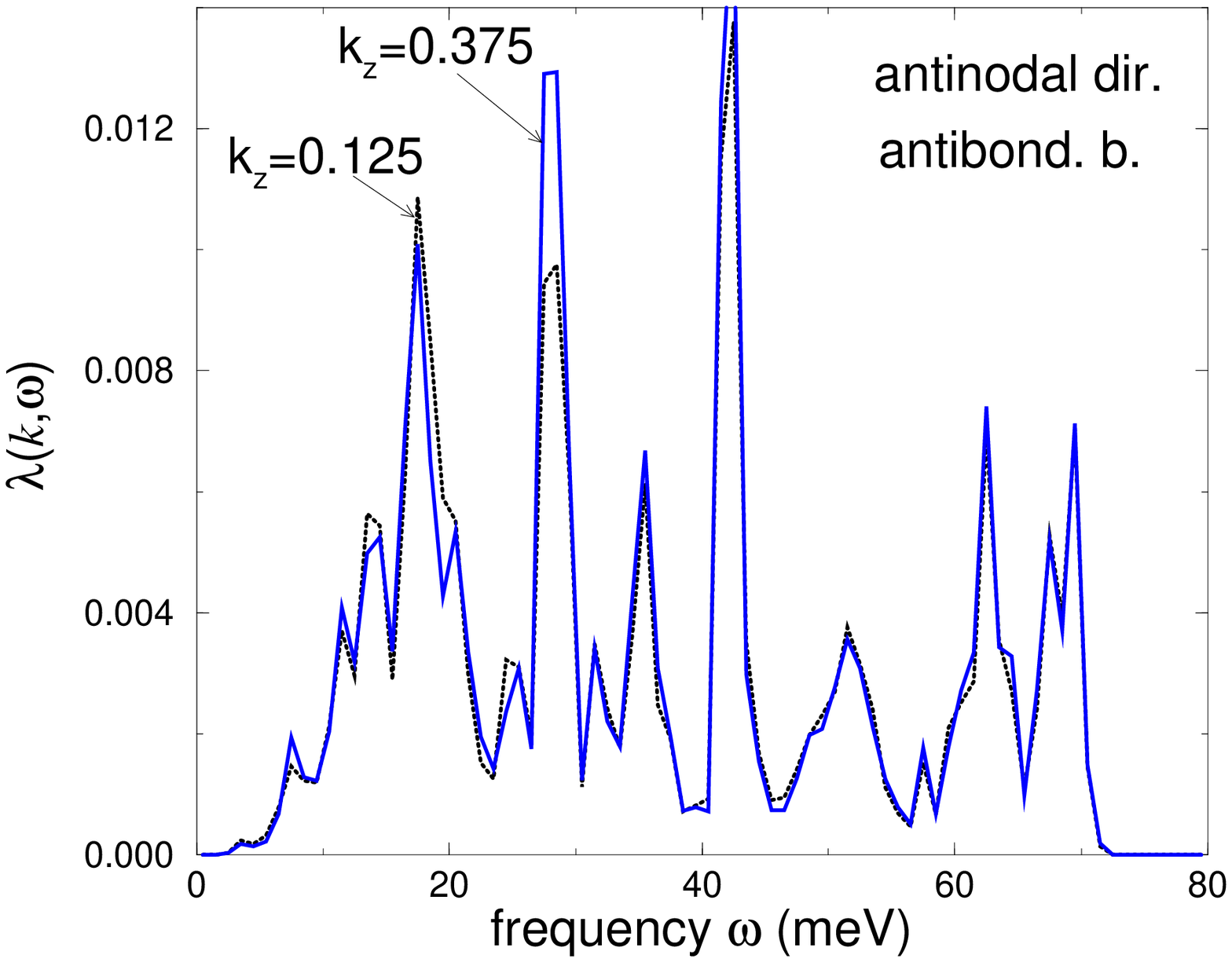}}
%\vspace{-0.5cm}
\caption{\label{fig:2}
(Color online)
Momentum and frequency resolved coupling function $\lambda({\bf k},\omega)$
in 1/meV of the antibonding band near the Fermi surface in the nodal 
(upper diagram) 
and the antinodal (lower diagram) direction. Solid and dashed lines 
correspond to $k_z=0.375$ and $0.125$, respectively. 
}
\end{figure}
%%%%%%%%%%%%%%%%%%%%%%%%%%%%%%%%%%%%%%%%%%%%%%%%%%%%%%%%%%%%%%%%%%%%%%%%%%%%

The following results refer to the antibonding band, thus we drop
from now on the band index $\nu$.
Fig.\ref{fig:2} shows frequency-resolved coupling functions 
$\lambda({\bf k},\omega)$ 
for momenta near the Fermi line in nodal (upper diagram) and
antinodal (lower diagram) directions. The solid lines correspond
to $k_z=0.375$, the dashed lines to $k_z=0.125$. The differences 
between solid and dotted lines are very small reflecting the 
very small $k_z$-dependence of $\lambda$. The spectra are very broad
and extend over the entire one-phonon range.
It is evident that models which use only  one or a few
selected phonons are not able to represent adequately the phonon spectrum.

The imaginary part of the retarded electronic self-energy due to phonons 
can be written as\cite{Mahan}
\begin{eqnarray}
Im \Sigma({\bf k},\omega)
&=&-\frac{\pi}{2} \int_0^\infty du u \lambda({\bf k},u)
\nonumber\\
&\Big[&\hspace*{-0.2cm}2b(u)+f(u-\omega)+f(u+\omega)\Big].
\label{Sigma}
\end{eqnarray}
Here, $b$ and $f$ denote the Bose and the Fermi function, respectively, 
and we assumed the self-energy to be diagonal in the band index.
The corresponding real part of $\Sigma$ was obtained by a
Kramers-Kronig transformation using a cutoff of 1 eV. 
%%%%%%%%%%%%%%%%%%%%%%%%%%%%%%%% FIG. 3 %%%%%%%%%%%%%%%%%%%%%%%%%%%%%%%%%%%%
\begin{figure}[htpb] 
\includegraphics[width=6.2cm]{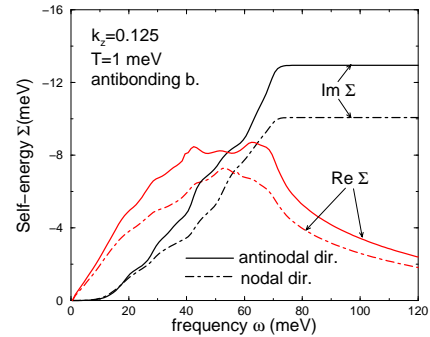}
\vspace{-0.5cm}
\caption{\label{fig:3}
(Color online)
Real and imaginary parts of the self-energy $\Sigma$ of the antibonding band
near the Fermi points along the nodal and antinodal directions at the 
temperature T=1 meV.
}
\end{figure}
%%%%%%%%%%%%%%%%%%%%%%%%%%%%%%%%%%%%%%%%%%%%%%%%%%%%%%%%%%%%%%%%%%%%%%%%%%%%#
Fig.\ref{fig:3} shows the 
calculated self-energy near the nodal and antinodal Fermi points  
using the coupling functions of Fig.\ref{fig:2}.
The negative imaginary part of $\Sigma$ increases monotonically over the
whole phonon spectrum and then approaches a constant. The negative real
part increases first, passes then through a maximum roughly in the 
middle of the phonon spectrum and then decreases monotonically.  
Both the real and imaginary parts of $\Sigma$ exhibit small oscillations
as a function of frequency which reflect the many maxima in 
$\lambda({\bf k},\omega)$ and the complexity of the phonon spectrum.
At elevated temperatures (for instance, at T=9 meV) these oscillations
vanish, the maximum in $Re \Sigma$ becomes somewhat smaller and 
$Im \Sigma$ finite at zero frequency but otherwise there is no
substantial change in $\Sigma$. Fig.\ref{fig:3} also illustrates the 
small anisotropy of $\Sigma$:
In the antinodal direction $\Sigma$ is only slightly larger 
compared to the nodal direction while the position of the maximum
of $Re \Sigma$ is approximately the same. For the nodal direction
experimental values for the maximum of $Re \Sigma$ are
20-40 meV\cite{Gromko,Zhou} whereas we have only about 7 meV at T=9 meV. 
%%%%%%%%%%%%%%%%%%%%%%%%%%%%%%%% FIG. 4 %%%%%%%%%%%%%%%%%%%%%%%%%%%%%%%%%%%%
\begin{figure}
\includegraphics[width=6.2cm]{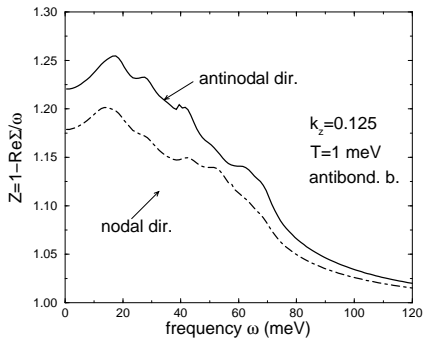}
\caption{\label{fig:4}
Renormalization function $Z({\bf k},\omega)$ of the antibonding band near 
the Fermi points in nodal and antinodal directions for $k_z=0.125$
at the temperature T=1 meV. 
}
\end{figure}
%%%%%%%%%%%%%%%%%%%%%%%%%%%%%%%%%%%%%%%%%%%%%%%%%%%%%%%%%%%%%%%%%%%%%%%%%%%%%

Fig.\ref{fig:4} shows the frequency dependence of the renormalization
function $Z$, defined by 
$Z({\bf k},\omega) = 1-Re\Sigma({\bf k},\omega)/\omega$,
near the nodal and antinodal Fermi points. At zero frequency $Z$
is equal to $1+\lambda({\bf k})$ where $\lambda({\bf k})$
is the frequency-integrated, dimensionless coupling constant depending
still on the momentum. This coupling constant is about 0.18 and 0.22
at the nodal and antinodal points, respectively, illustrating again the 
smallness
of the electron-phonon coupling. With increasing frequency $Z$
increases first slightly, passes through a maximum and then approaches
unity beyond the end of the one-phonon spectrum.
Superimposed on this general behavior are many small oscillations
which reflect the numerous density peaks in Fig.\ref{fig:2} and which
vanish already at T=9 meV. 
If, as in our case, the imaginary part of $\Sigma$ is rather small,
the dispersion of the electrons is given by the equation
$\omega = \epsilon_{\bf k}/Z({\bf k},\omega)$ which for small
$\Sigma$'s has the solution  
$\omega = \epsilon_{\bf k}/Z({\bf k},\epsilon_{\bf k})$.
A kink in the dispersion is generated if $Z$ changes from
a large to a small value in a narrow frequency interval. This happens,
for instance, in a model with one single optical phonon, see Fig.2 of  
Ref.\cite{Zeyher}. However, for our realistic phonon spectrum and coupling 
functions
this does not occur according to Fig.\ref{fig:4}, so there will be 
no well-pronounced kink feature in the electronic dispersion. There seems 
to be a general
consensus that ARPES data for the nodal and antinodal directions
yield kink energies of about 70 and 40 meV, respectively. 
It is clear that these energies do not represent any characteristic energies in
the frequency dependence of $Z$ in Fig.\ref{fig:4}.

\begin{figure}[ttt] 
\vspace*{-12ex}
\includegraphics[width=8.2cm]{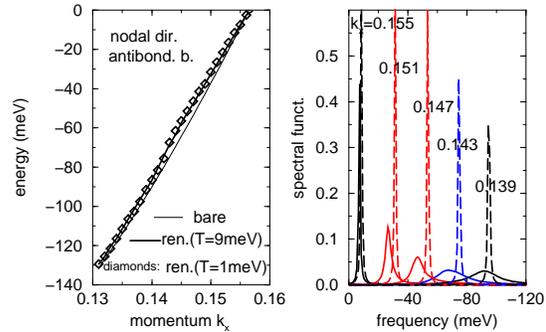}
\vspace{-0.5cm}
\caption{\label{fig:5}
(Color online)
Left: Electronic dispersion; right: spectral function 
of the antibonding band for different momenta $k_x=k_y$ along the nodal 
direction using T=1 meV and $k_z=0.125$.  
}
\end{figure}

Finally, the left diagram in Fig.\ref{fig:5} illustrates
the weakness and smoothness of the renormalization of the 
dispersion of the antibonding band in the nodal direction and also
its weak dependence on temperature.
The ratio of asymptotic slopes at small and large frequencies is about 1.18
which is much smaller than typical experimental values of about 2 in optimally 
doped cuprates\cite{Lanzara,Gromko,Koralek}. 
The right diagram in Fig.\ref{fig:5}
shows spectral functions of the antibonding band  
along the nodal direction for five different momenta $k_x=k_y$. 
Unlike in the case of one optical phonon coupled strongly to electrons,
discussed in Ref.\cite{Zeyher}, the solid lines show always only one peak which
departs and then returns very smoothly from the noninteracting 
peak with decreasing energy.

Based on the above LDA results our main conclusions are: a) The 
coupling function $\lambda({\bf k},\omega)$ shows as a
function of ${\bf k}$ a minimum at the nodal and a maximum
at the antinodal Fermi point rather independently of the involved phonon
modes. The resulting anisotropy in $\bf k$ is rather moderate both for the 
bonding and the antibonding band;
b) It is important to keep a realistic phonon spectrum and not only one
or two phonon branches because the $\omega$-dependence of $\lambda$
is mainly determined by phonon densities where, for instance, the breathing
or the buckling mode do not play any distinguished role; c) The
frequency-integrated coupling function $\lambda({\bf k})$ varies between
0.18 and 0.22 along the Fermi surface. Due to these rather small values
the phonon-induced electronic self-energies
along the nodal direction are too small by about a factor 3-5 to account for 
the self-energies deduced from ARPES experiments and for the observed slope 
changes in the electronic dispersion, at least, within the LDA.
%%%%%%%%%%%%%%%%%%%%%%%%%%%%%%%% FIG. 5 %%%%%%%%%%%%%%%%%%%%%%%%%%%%%%%%%%%%
%%%%%%%%%%%%%%%%%%%%%%%%%%%%%%%%%%%%%%%%%%%%%%%%%%%%%%%%%%%%%%%%%%%%%%%%%%%%%

The authors thank T.P. Devereaux for discussions and clarifications
concerning, in particular, the correct definition of $\lambda$ in a 
multi-band superconductor.
%%%%%%%%%%%%%%%%%%%%%%%%%%%%%%%%%%%%%%%%%%%%%%%%%%%%%%%%%%%%%%%%%%%%%%%%%%%%
%%
%%                           REFERENCES
%%
%%%%%%%%%%%%%%%%%%%%%%%%%%%%%%%%%%%%%%%%%%%%%%%%%%%%%%%%%%%%%%%%%%%%%%%%%%%%

\end{document}